\documentclass[12pt,preprint]{aastex}

\newcommand{\lta}{\lesssim}

\newcommand{\cotwo}{$\rm ^{12}CO(2-1)$}

\newcommand{\ione}{$I_{10}$}
\newcommand{\itwo}{$I_{21}$}

\newcommand{\solm}{M_{\odot}}
\newcommand{\solms}{\>{\rm M_{\odot}/pc^2}}

\newcommand{\ea}{et al.}
\newcommand{\kkms}{\>{\rm K}\,{\rm km}\,{\rm s}^{-1}}
\newcommand{\kms}{\>{\rm km}\,{\rm s}^{-1}}
\newcommand{\gyr}{\>{\rm Gyr}}
\newcommand{\myr}{\>{\rm Myr}}
\newcommand{\pc}{\>{\rm pc}}
\newcommand{\kpc}{\>{\rm kpc}}
\newcommand{\mpc}{\>{\rm Mpc}}
\newcommand{\m}{\>{\rm m}}

\newcommand{\mjy}{\>{\rm mJy}}
\newcommand{\msun}{\>{\rm M_{\odot}}}

\newcommand{\dg}{^{\circ}}

\newcommand{\as}{^{\prime\prime}}

\newcommand{\bdm}{\begin{displaymath}}
\newcommand{\edm}{\end{displaymath}}
\newcommand{\beq}{\begin{equation}}
\newcommand{\eeq}{\end{equation}}
\newcommand{\bit}{\begin{itemize}}
\newcommand{\eit}{\end{itemize}}
\newcommand{\ben}{\begin{enumerate}}
\newcommand{\een}{\end{enumerate}}
\newcommand{\bfi}{\begin{figure}[htb]}
\newcommand{\bpfi}{\begin{figure}[p]}

\newcommand{\htwo}{$\rm H_2$}

\newcommand{\ha}{$\rm H\alpha$}
\newcommand{\mhtwo}{$M_{\rm H_2}$}

\shorttitle{CO(2-1) at the Nucleus of IC342}
\shortauthors{Schinnerer, B\"oker \& Meier}

\begin{document}




\def\xx{$^{[xx]}$}

\def\tbd#1{{\baselineskip=9pt\medskip\hrule{\small\tt #1}
\smallskip\hrule\medskip}}

\def\com#1{{\baselineskip=9pt\medskip\hrule{\small\sl #1}
\smallskip\hrule\medskip}}

\title{Molecular Gas and the Nuclear Star Cluster in IC342:\\
Sufficient Inflow for Recurring Star Formation Events?}

\author{Eva Schinnerer\altaffilmark{1,4}, Torsten
B\"oker\altaffilmark{2,5} \& David S. Meier\altaffilmark{3} }
\altaffiltext{1}{National Radio Astronomy Observatory, P.O. Box 0, 
                 Socorro, NM 87801}
\altaffiltext{2}{Space Telescope Science Institute, 3700 San Martin Drive, 
	Baltimore, MD 21218, U.S.A.}
\altaffiltext{3}{University of Illinois, Urbana-Champaign, 1002 W. Green St., 
	Urbana, IL 61801}
\altaffiltext{4}{Jansky-Fellow}
\altaffiltext{5}{On assignment from the Space Telescope Division
	of the European Space Agency (ESA).}


\begin{abstract}       
We present high spatial resolution ($1.2\as$) mm-interferometric
observations of the \cotwo\ line emission in the central $300\pc$ of
the late-type spiral galaxy IC~342. The data, obtained with the Owens
Valley Radio Observatory, allow first-time detection of a molecular
gas disk that coincides with the luminous young stellar cluster in the
nucleus of IC\,342. The nuclear CO disk has a diameter of $\sim 30\pc$
and a molecular gas mass of $\rm M_{H_2} \sim 2\times10^5\msun$. It
connects via two faint CO bridges to the well-known, $100\pc$ diameter
circumnuclear gas ring. Analysis of the gas kinematics shows that the
line-of-nodes in the inner $15\pc$ is offset by about $45\dg$ from the
major kinematic axis, indicating non-circular motion of the gas within
a few parsec of the dynamical center of IC\,342. Both the morphology
and the kinematics of the CO gas indicate ongoing inflow of molecular
gas into the central few parsec of IC\,342.
We infer a gas inflow rate between 0.003 and $\rm 0.14\>\msun yr^{-1}$, 
based on the observed surface density of the nuclear gas disk
and estimates of the radial velocities of the surrounding
gas. Inflow rates of this order can support repetitive star
formation events in the nucleus of IC\,342 on timescales much smaller
than a Hubble time.
\end{abstract}

\section{INTRODUCTION}
Recent observations with the Hubble Space Telescope (HST) at both
optical and near-infrared wavelengths \citep[e.g.][]{car98,boe02} have
revealed that the very centers of late-type spiral galaxies are often
occupied by a photometrically distinct, luminous, and compact (with a
few parsec diameter) stellar cluster. Most of the nuclear star
clusters studied in detail appear to have experienced recent star
formation. For example, M\,33 \citep{gor99}, NGC\,4449 \citep{boe01},
as well as NGC\,247 and NGC\,2403 \citep{dav02} all have nuclear
clusters with a spectral energy distribution (SED) that is dominated
by a young ($\lta 100\myr$) stellar population. Clearly, these
galaxies must recently have experienced significant inflow of
molecular gas into their central few parsec in order to trigger the
formation of these young clusters. There is mounting evidence that
large-scale bars can efficiently move large quantities of molecular
gas down to the inner kiloparsec \citep{sak99}. However, the exact
mechanism which can transport gas from a few hundred parsec down to a
few parsec is still under debate.

IC~342 is a prime example of a late-type spiral galaxy with a young
luminous nuclear cluster which formed in a short-lived burst about
$60\myr$ ago \citep*{boe97,boe99}. For this paper, we adopt the
distance of $1.8\mpc$ derived by \cite{mcc89} which makes IC\,342 the
closest late-type galaxy with strong nuclear star
formation\footnote{The distance to IC~342 is uncertain, e.g. more
recent observations indicate a larger distance of 3.3\,Mpc
\citep*{sah02}. This should be kept in mind when discussing
distance-dependent quantities.}. It thus offers a unique opportunity
to study the related gas dynamics on scales of a few parsec ($1\as\sim
8.7\pc$). The central region of IC~342 has been mapped in the past in
several molecular transitions on scales of a few arcseconds
\citep[e.g.][]{lo84,ish90,tur92,tur93,mei00,mei01}. These maps show
that the molecular gas is located in two spiral arms which show strong
streaming motions. The spiral arms join to form a ring of about
$10\as$ diameter which is the site of massive star formation
\citep{boe97}. In this Letter, we present a high resolution, high
sensitivity \cotwo\ map which for the first time resolves the
molecular gas kinematics {\it inside} the starburst ring.
\section{OBSERVATIONS}
The center of IC~342 was observed in the \cotwo\ line
in January 2002 with the six-element Owens Valley Radio Observatory
(OVRO) millimeter interferometer. Obtained in the L and H configurations with
baselines between 15 and $115\m$, the data have a spatial resolution
of $\sim 1.2\as$ ($10\pc$) with robust weighting. The noise per
$2.601\kms$ wide channel is $\rm 40\mjy$ per beam in the combined data of both
tracks. For the intensity map, we used only emission that is $5\sigma$
above the noise floor and evident in at least two adjacent channels. Similarly,
we used a $15\sigma$ blanking for the velocity field to emphasize the
small-scale velocity structure. Nevertheless, our data recover most of
the CO flux in the region mapped, as the peak integrated intensity
of our data convolved to $14\as$ resolution is about 90\% of the value
of $\rm I_{30m}=324\kkms$ measured with the IRAM 30m single dish
telescope \citep*{eck90}. The absolute astrometry of the OVRO
observations are tied to the VLA quasar reference frame, which are
known to $\leq 0.02''$. However, the positional uncertainty in the 
observations is limited by atmospheric refraction at 230 GHz and 
is $< 0.1''$.

For comparison to the CO data, we retrieved {\it Hubble Space
Telescope} V-band (F555W) and \ha\ (+continuum, F658N) maps 
from the newly available WFPC2 associations
archive. In order to register the HST images and the CO maps, we
aligned the eastern CO spiral arm to the prominent dust lane seen in
both HST images. Incidentally, this also aligned the central CO peak
with the nuclear star cluster. The alignment required a nominal shift
of $1.28\as$ to the west and $0.59\as$ to the north\footnote{The HST
coordinates were measured from the positions of 8 stars in the field
of view that are listed in the USNO-A2.0 catalog.}. Potential
discrepancies of this magnitude between HST and radio coordinate
systems are not uncommon \citep[e.g.][]{whi02}.
\section{MOLECULAR GAS PROPERTIES\label{sec:data}}
\begin{figure}
\epsscale{0.8}
\rotatebox{-90}{
\plotone{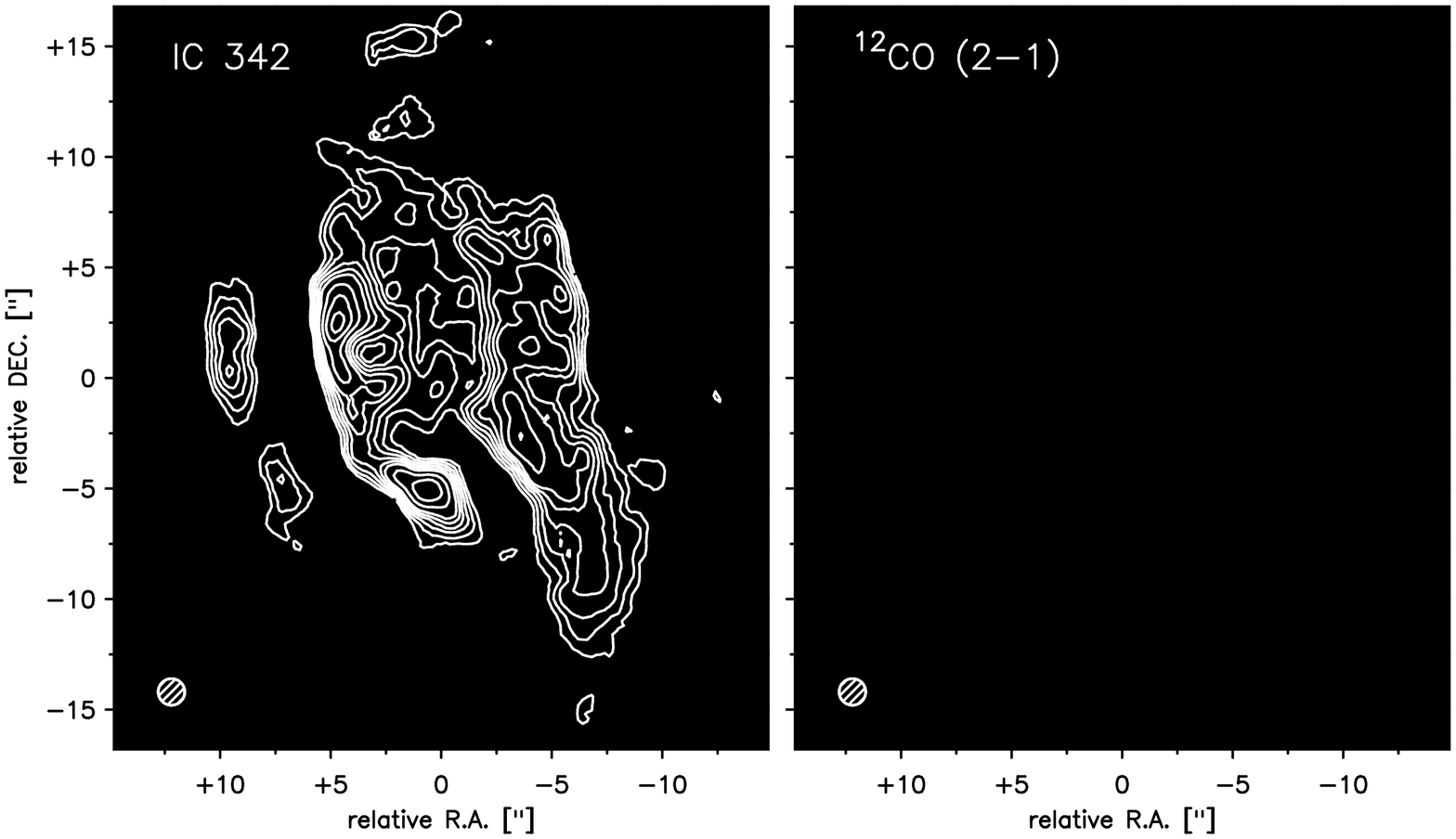}}
\caption{
OVRO \cotwo\ intensity map ({\it left}) and velocity field ({\it right})
at $10\pc$ ($1.2\as$) resolution. The contours in the intensity map are
at 4, 6, 8, 10, 12, 14, 18, 22, and $26\sigma$ ($1\sigma = 
2.6\,{\rm Jy\,beam^{-1}\,\kms}$). 
The coordinate center is at 03:46:48.26, +68:05:47.8 (J2000). 
The beam size is shown in the lower left corner.
\label{fig:co}}
\end{figure}

%
%
{\bf CO Distribution and Geometry:}
The intensity map in Figure~\ref{fig:co} partly resolves the structure 
of the two well-known molecular spiral arms 
\citep[e.g.][]{lo84,tur92,mei00,mei01}. The two arms have quite different 
morphologies: the eastern arm forms a continuous spiral which is
barely resolved with our data. A second, weaker component about $5\as$
east of the main arm is also apparent in the map. The western arm, on
the other hand, has two main components: a southern part pointing
toward the nucleus, and a northern part (north of $-3\as ,0\as$) which
seems to form another spiral. The western arm is about $5\as$ wide and
has a sharp inner edge, but becomes more diffuse toward larger
radii. The mean gas density within the spiral arms is $\rm \sim
10^{23}\>cm^{-2}$, assuming an intrinsic line ratio of \itwo /\ione =
1 and the standard Galactic conversion factor of $\rm
2\times10^{20}\,cm^{-2}\,(K\,km\,s^{-1})^{-1}$ \citep{str88}. These
values translate into an average gas surface densities of $\sim
1100\,\solms$ (including a 36\% contribution from He), with peak
values about twice as high. The map reveals somewhat weaker emission
inside the central trough that has been identified for the first
time. Because this emission component has a disk-like morphology and
appears centered on the nuclear star cluster (see below), we refer to
it as the {\it nuclear disk} for the remainder of this paper. The disk
has a major axis diameter of $\sim 30\pc$ and a gas mass of $\sim
1.7\times10^5\msun$, with a mean gas density about half of that in the
spiral arms. The disk appears connected to both spiral arms via two
faint CO ``bridges'' $\sim 1\as$ north and south of the nucleus.
%
%
\\
\\
{\bf CO Kinematics:}
The overall motion of the molecular gas is from southwest to 
northeast, in agreement with the position angle of the kinematic major
axis of $\sim 37\dg$ determined from HI data \citep{cro01}. 
The arms of the gas spiral show strong streaming motions, and
are interpreted as gas lanes along the leading side of a weak stellar
bar \citep{tur92,sak99}. On smaller scales, the CO velocity
field within the spiral arms (Fig.~\ref{fig:co}) does not appear well 
ordered since the line emission consists of multiple components. 
The high spatial and spectral resolution
of the OVRO data allows us to identify non-circular gas motions close
to the center of IC~342. The molecular gas is moving along a position
angle of $\sim -10\dg$ (Figs.~\ref{fig:co} and \ref{fig:renzo}) which
is offset by $\sim 45\dg$ from the kinematic major axis of the galaxy
disk. This is the first time that non-circular
motions within $10\pc$ of the nucleus have been detected in IC~342,
their nature is discussed further in section \ref{sec:feed}. 
Within the spatial resolution of our observations, the central CO peak
coincides with the dynamical center as well as the nuclear star
cluster, and the systemic velocity we measure is $v_{\rm LSR}\approx
46.0\,\kms$. This value is $\sim 7\,\kms$ higher than inferred from
lower resolution CO(1-0) data \citep[$\sim 39\kms$][]{sak99}. This
discrepancy is likely due to the non-circular nature of the nuclear CO
kinematics which only becomes more apparent at high angular
resolution.
%
%
\\
\\
{\bf Comparison to Optical Morphology:}
The distribution of the molecular gas differs from that of the stellar
inventory as demonstrated by the comparison of the HST images to the
CO distribution (Fig. \ref{fig:hst}). The CO spiral arms outline the
bright central emission region that is evident in both the H$\alpha$
emission and the V-band continuum. Both the line emission and the
optical continuum are brighter east of the nucleus where the CO
emission is confined to the prominent dust lane. In addition, optical
light fills the gap between the two eastern gas lanes. On the other
hand, west of the nucleus where the CO spiral is much wider, there is
a lack of H$\alpha$ emission and optical continuum. This asymmetry in
the extinction pattern together with the velocity field suggests that
the western side is the near side of IC~342. A prominent H$\alpha$
knot lies on the boundary (-$3\as$;-$1\as$) between the two parts of
the western gas spiral, it coincides with Region~1 studied by
\cite{boe97} and is the site of intense star formation. Within the
nuclear disk, the peak of the molecular gas (which is close to the
dynamical center, see \S~\ref{sec:data}) coincides roughly
with the nuclear star cluster. Two other star clusters north
and south of the nucleus fall into regions inside the spiral arms
which are almost devoid of molecular line emission. The HST V-band
image shows two fine dust lanes of $\sim 0.2\as$ width about $1\as$
north and south of the nucleus which extend in east-west direction
(best seen in Fig. \ref{fig:renzo}). The two CO bridges that connect
the nuclear disk to the spiral arms appear to coincide with these dust
lanes, although the lower spatial resolution of the CO map makes this
identification somewhat tentative.
\section{DISCUSSION\label{sec:feed}}
\begin{figure}
\epsscale{0.9}
\plotone{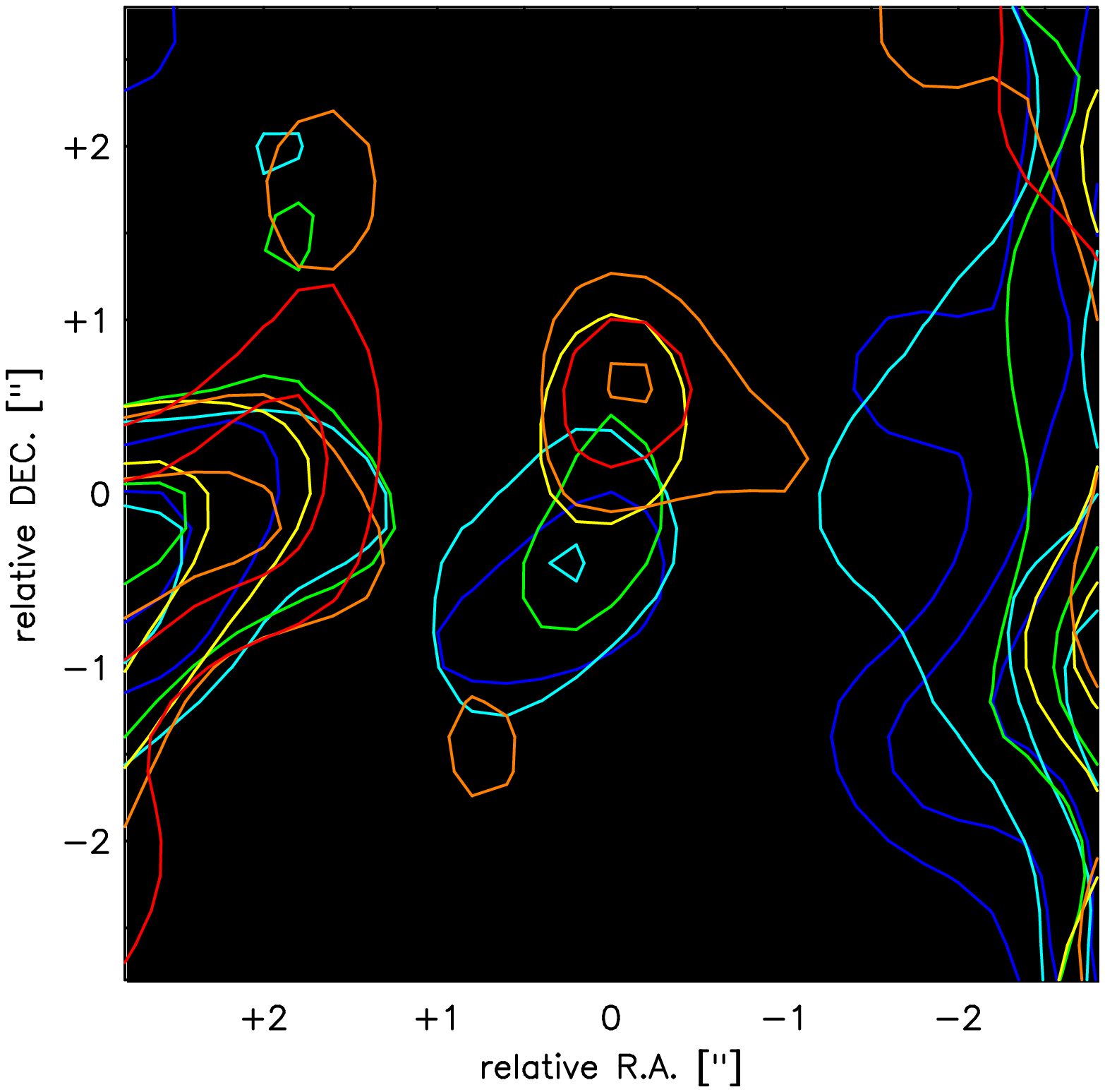}
\caption{
Renzogram of \cotwo\ emission. Close to the
nucleus, the gas is moving along a position angle of $\sim -10\dg$. The
contours represent different channel maps from
$41.523\kms$ ({\it blue}) to $54.528\kms$ ({\it red}) in steps of
$2.601\kms$. The contours for each channel map are at $15\sigma$, 
$20\sigma$ and $25\sigma$ with $\sigma$=40\,mJy per beam. 
The HST F555W image is shown in gray-scale.
\label{fig:renzo}}
\end{figure}

\begin{figure}
\epsscale{1.05}
\rotatebox{-0}{
\plotone{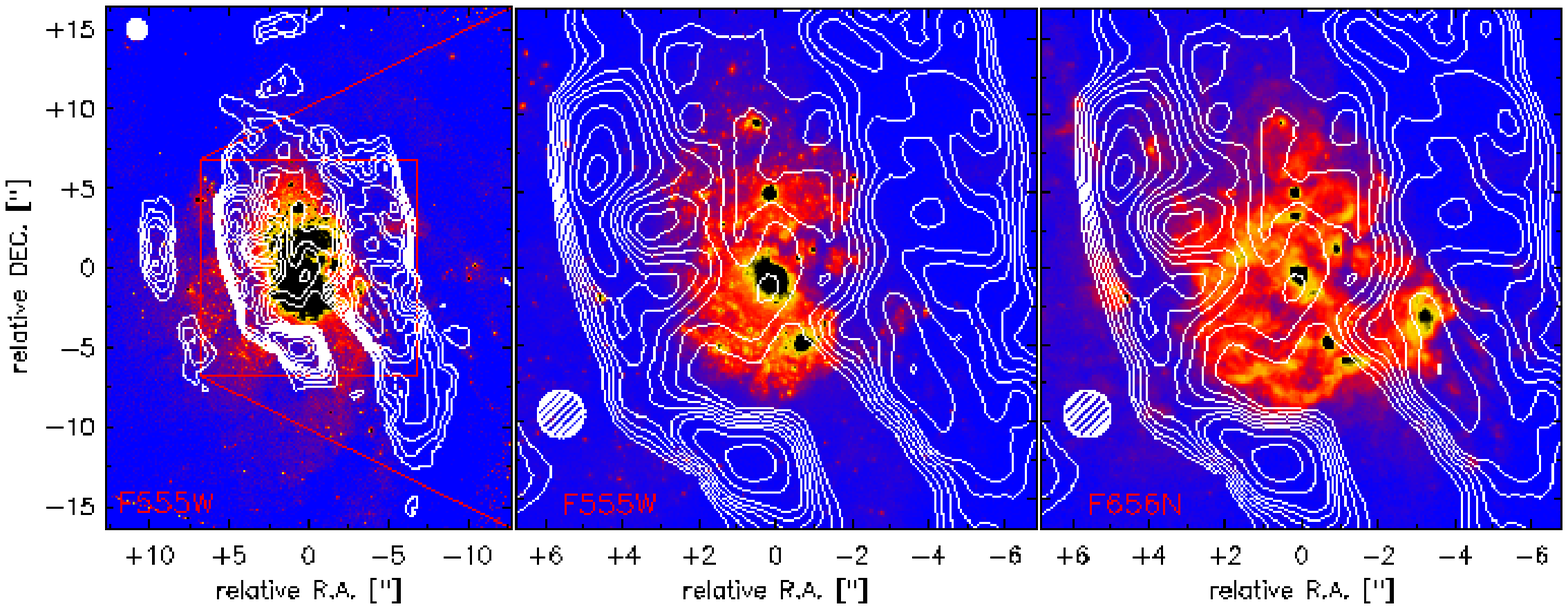}}
\caption{
The OVRO \cotwo\ intensity map overlaid onto the archival HST V-band 
({\it left, middle}) and \ha\ image ({\it right}). The left panel shows
the full OVRO field, while the middle and right panels contain only 
the inner $120\pc$ (indicated by a red box in the left panel). Different
color stretches are used to enhance low level emission ({\it left}) 
as well as distinct features in the nuclear region ({\it middle}) 
of the HST F555W image. The OVRO CO contours are the same as
in Figure~\ref{fig:co}. For reference, the beam size is shown in each
panel.
\label{fig:hst}}
\end{figure}

{\bf What Drives the Gas Inflow?}
The overall shape and kinematics of the CO spiral arms are reminiscent
of models for gas flows in a weakly barred potential
(e.g. Athanassoula 1992). Although no stellar bar is obvious in the
optical data, ellipse fits to deep wide-field CCD images show evidence
for a bar with $600-700\as$ length and a position angle of $\sim
20\dg$ \citep{but99}. Since the western side is the near side (see
\S~\ref{sec:data}), the sense of rotation is counter-clockwise, and
therefore the gas spiral arms are on the leading side of this
large-scale bar. \cite{cro01} interpret the displacement of the
optical and dynamical position angles and the warped outer HI disk as
signs of a recent mild interaction. This interaction might have
triggered the formation of the oval distortion which in turn might be
responsible for the pile-up of molecular gas in the central kiloparsec
of IC\,342. The molecular gas morphology of IC~342 is quite similar to
the one observed in the center of our own Galaxy where a 
$\sim 14\pc$ wide molecular disk or torus is surrounded by a 
molecular ring with a radius of $180\pc$
\citep[see review by][]{mor96}. Also, the bar length of $\sim
5.2\kpc$ in IC~342 is comparable to the $4.8\kpc$ observed in our
Galaxy. The filamentary structure of the dust/gas lanes and the
H$\alpha$ emission in IC~342 (similar filaments are also observed in
the inner $500\pc$ of our Galaxy) might be explained by 
self-gravitating gas \citep[see Fig. 2 of][]{wad01}.
The coincidence of the CO disk's central velocity with systemic
velocity favors a scenario in which the gas lies within the disk, as
opposed to being situated above the plane. The in-plane picture is
supported by the good agreement between the gas distribution and the
dust morphology.
\\
In the following we will discuss three plausible interpretations of
the CO morphology. First, the nuclear CO disk could be tilted relative
to the large stellar disk. The axis ratio of the nuclear disk of $b/a
\sim 0.5$ translates into an inclination of $i \sim 60\dg$ along a
position angle of $\sim 15\dg$. The two CO peaks about $0.5\as$ north
and south of the center are separated in velocity by $8\kms$ (see
Fig. \ref{fig:renzo}). Assuming solid body rotation and that these two
peaks fall on the kinematic major axis of the nuclear disk, the
resulting dynamical mass is $M_{\rm dyn} \sim 2\times10^4\msun$. The
derived mass is significantly smaller than the mass estimated for
either the nuclear molecular gas disk or the nuclear stellar cluster
\citep[$6\times10^6\msun$,][]{boe99} making this an unlikely
scenario. A second alternative is that the nuclear CO emission
originates in a virialized cloud, rather than a rotating disk. For a
cloud size of about $2\as (17\pc)$ and a CO line width (FWHM) of $\Delta v
\sim 8\kms$, the virialized mass is $M_{\rm vir}
\sim 3\times10^5\msun$. However, the observed rotation (see
Fig. \ref{fig:renzo}) within the nuclear disk argues against a single
virialized cloud. In a third scenario, the observed non-circular
motion are due to radial motions of material moving toward the center
similar to what is observed in the central region of our Galaxy
\citep*{mor96}. Since the first two scenarios seem unlikely, we
conclude that streaming motions are the most plausible explanation for
the non-circular motions observed in the CO kinematics. A possible
cause for such streaming motion could be the large-scale stellar
bar. This scenario should be investigated further by dynamical modeling.
\\
\\
{\bf Feeding the nuclear star cluster:}
The nuclear star cluster in IC\,342 has a mass of $6\times10^6\msun$
and experienced a major burst of star formation $\sim 60\myr$ ago
\citep{boe99}. The population of stars produced in this event
dominates the SED of the cluster, and thus makes it difficult to
quantify the contribution from possible older generation(s) of
stars. As discussed in \cite{boe99}, up to 90\% of the cluster mass
could have been produced in a burst that occurred $10\gyr$ before the
most recent one.  While this number is definitely an upper limit
because the described scenario lies at the extreme end of possible
star formation histories, it is reasonable to assume that previous
nuclear starburst events have indeed occurred.  For the moment, we will
assume that the most recent burst produced 30\% of the cluster mass,
i.e. about $2\times10^6\msun$.  Assuming a star formation efficiency
of 20\% \citep*{elm99}, about $10^7\msun$ of molecular gas are needed
in order to produce such a starburst event.  This is 50 times more
than the present-day mass of the nuclear CO disk (\mhtwo $\sim
1.7\times 10^5\msun$). In what follows, we use our CO observations to
constrain the inflow rate of molecular gas and to estimate how long it
might take before another such starburst event can occur.
\\
A lower limit can be calculated by assuming that it took the full
$60\myr$ since the last starburst for the nuclear CO disk to form.  In
this case, the inflow rate is $\sim 0.003\msun/{\rm yr}$ or about 3
Gyr are needed before a new starburst event could occur. We derive an
upper limit by taking the observed gas surface density of $\Sigma \sim
500\solms$ together with the width of the two dust lanes seen in the
V-band image ($\sim 0.2\as$). This yields an estimate of the gas inflow rate
onto the nuclear disk of $\sim 2000\msun /\pc
\times v_{\rm rad}$. The inflow velocity of the gas, $v_{\rm rad}$ 
can be estimated by comparing the observed circular velocity in the
nuclear disk ($v_{\rm obs} \approx 25\kms$ at a radius of 
$0.5\as$ or $4.4\pc$) to the maximum expected circular velocity around
the nuclear star cluster (assuming solid body rotation), $v_{\rm sb}\approx
75\kms$. This implies that radial motions of up to $v_{\rm rad}
\sim 70\kms$ can be present in the molecular gas next to the nuclear
cluster, depending on the exact shape of the gravitational
potential. These rough estimates result in an inflow rate of $\rm \sim
0.14\,\solm/yr$. Such an inflow rate would accumulate about
$10^7\msun$ of molecular gas in the nucleus within $70\myr$.
These two estimates probably bracket the actual inflow rate, but a
number of uncertainties remain. For one, the \htwo\ mass in the center
of IC~342 might be overestimated by a factor of 4-5 due to the use of
a Galactic conversion factor, as discussed in \cite{mei01}. In
addition, the non-circular motions can be overestimated due to the
unknown shape of the underlying gravitational potential.
Nevertheless, the inflow rates discussed here are not unreasonable to
support repetitive nuclear starbursts in IC\,342.
%
For example, an accumulation of $\sim 0.01\msun/{\rm yr}$ is
sufficient to support a starburst duty cycle of about $1\gyr$. The
duty cycle might even be shorter if the star formation efficiency is
higher than the 20\% assumed here. Indeed, efficiencies up to 50\%
have been inferred for giant molecular clouds in the spiral arms of
IC~342 \citep{mei01}. Moreover, the inflow rate is likely to vary over
time with lower inflow rates in the past, and could increase
in the future as more mass is accumulated at the center. For example,
assuming a constant rate of $0.003\msun/{\rm yr}$ (our lower limit)
over a Hubble time ($10^{10}{\rm yr}$) results in a mass accumulation
of $\sim 3\times10^7\,\solm$ which is larger than the dynamical mass
of $\sim 2\times10^7\,\solm$ derived for the central 120\,pc
\citep{tur92}. 
\\
Our analysis of new high-resolution maps of the \cotwo\ emission in the
center of IC\,342 has revealed the following main results. We detect
molecular gas coinciding with the nuclear stellar cluster. This, and
the presence of streaming motions close to the nucleus, suggests that
gaseous matter is currently accumulating at the nucleus of
IC\,342. These findings provide support for in-situ formation of
nuclear clusters and repetitive nuclear starbursts. Rough estimates of
gas inflow rates suggest that massive ($10^6\msun$) nuclear starbursts
in IC\,342 can be supported with duty cycles between a few hundred
Myrs and 1 Gyrs. 

\acknowledgments
We thank J. Turner and W. Maciejewski for helpful comments. DSM
acknowledges support from the Laboratory of Astronomical Imaging at
the University of Illinois.

\clearpage

\end{document}